\documentclass[hyperref]{article}
\usepackage{makeidx}
\usepackage[colorlinks=true,citecolor=black]{hyperref}
\usepackage{hyperref}
\usepackage{multirow}
\usepackage{multicol}
\usepackage[dvipsnames,svgnames,table]{xcolor}
\usepackage{graphicx}
\usepackage{epstopdf}
\usepackage{ulem}
\usepackage{hyperref}
\usepackage{amsmath}
\usepackage{amssymb}
\usepackage{enumerate}
\usepackage[a4paper]{geometry}
\usepackage{subfigure}
\usepackage{caption}
\usepackage{tabulary}
\usepackage{makecell}
\usepackage{framed}
\makeatother 
\hypersetup{
colorlinks=true,
linkcolor=black
}
\begin{document}
\captionsetup[figure]{labelfont={bf},name={Fig.},labelsep=period}

\title{\bf Model Checking Logical Actions in Magic Tricks }

\date{}
\author{\large  Weijun ZHU\\
{\sffamily\small School of Computer and Artificial Intelligence, Zhengzhou University, Zhengzhou, 450001, China}
}
\maketitle
{\noindent\small{\bf Abstract:}
Some Magic Tricks (MT), such as many kinds of Card Magic (CM), consisting of human computational or logical actions. How to ensure the logical correctness of these MTs? In this paper, the Model Checking (MC) technique is employed to study a typical CM via a case study. First, computational operations of a CM called "shousuigongcishi" can be described by a Magic Algorithm (MAR). Second, the logical correctness is portrayed by a temporal logic formula. On the basis of it, this MT logical correctness problem is reduced to the model checking problem. As a result, the Magic Trick Model Checking (MTMC) technique aims to verify whether a designed MT meets its architect's anticipation and requirements, or not, in terms of logic and computations.}

\vspace{1ex}

\section{Introduction}
On a famous magic show on TV, Mr. Qian Liu, who is a well-known magician, showed his interesting CM, called "shousuigongcishi"[1]. A wonderful story, theatrics, perfect interaction, and other factors, enhanced the charm of this MT show. However, its key is a series of designed computational actions and a magic algorithm dealing with cards. The algorithm can be executed by a sole magician, and it can also be executed parallelly by countless television audience. In short, they were persons instead of computers, execute a magic algorithm, at that time.

The question is: are there logical bugs in the computational actions of the designed CM? Can we ensure that a magic algorithm, such as "shousuigongcishi", will never encounter unexpected situations at any time and in any state? According to media reports, a great number of viewers had already watched Mr. Liu's show [2], and the magician called on everyone to follow him and do it together on scene [1]. It is clear that, the designer must guarantee the logical correctness of this CM in advance, so that the use of formal verification and model checking are sound, in this case.

\section{MTM, Automaton and Algorithm for the Magician's Actions} 
{\bf Definition 1}, a finite word in an alphabet is defined as $\overline{a} = {a_1}{a_2}...$ \\
{\bf Definition 2}, a transition table is a quadruple $(\Sigma,S,{S_0},E) $.
\begin{itemize}
\item $\Sigma$ is a finite alphabet.
\item $S$ is a finite state set.
\item ${S_0} \subseteq S$ is a start state set.
\item $E \subseteq S \times S \times \Sigma$ is a set of transition rules, where $<s \times s \times a> \in E$ is a state transition.
\end{itemize}
{\bf Definition 3}, a finite run of a transition table $(\Sigma,S,{S_0},E) $ on a word $\overline{a}$ is a state sequance.\\
$r: {s_0} \xrightarrow{{a_1}} {s_1} \xrightarrow{{a_2}} ...... \xrightarrow{{a_n}} {s_n}$ \\
{\bf Definition 4}, a finite state automaton A is a quintuple $(\Sigma,S,{S_0},E,F) $, where $(\Sigma,S,{S_0},E) $ is a transition table, and $F$ is a set of final states. \\
{\bf Definition 5}, a finite state automaton $(\Sigma,S,{S_0},E,F) $ runs on a finite word, and this running is accepted if and only if \\
$r: {s_0} \xrightarrow{{a_1}} {s_1} \xrightarrow{{a_2}} ...... \xrightarrow{{a_n}} {s_n}$ and $\overline{a} = {a_1}{a_2}...{a_n}$ and ${s_n} \in F$ \\
{\bf Definition 6}, a regular language $L(A)$ is accepted by a finite state automaton $A$ if and only if $L(A)= \lbrace \overline{a} | there \, are \, accepting \, running \, r \, on \, \overline{a} \, in \, A \rbrace $ \\
{\bf Definition 7}, if each word $a$ means a magic action, a finite state automaton $A$ which is defined by from definition 1 to definition 5, is called a magic automaton. \\
{\bf Definition 8}, a finite running Turing machine can be specified as a quadruple $A=(Q,\Sigma,s,\delta) $ [3], where
\begin{itemize}
\item $\Sigma$ is a finite alphabet.
\item $Q$ is a finite state set.
\item $s$ is the initial state $s \in Q$
\item $\delta$ is a transition function determining the next move: $\delta : ( Q \times \Sigma ) \to ( \Sigma \times \lbrace L,R \rbrace \times Q ) $
\end{itemize}
{\bf Definition 9}, If each action of a Turing machine is a magic action, such a Turing machine is a magic Turing machine. \\
Obviously, the formal model portraying sousuigongcishi has a stronger expressive power than an automaton, so that sousuigongcishi can be modeled by a MTM instead of a magic automaton. Thus, a magic algorithm is proposed for describing sousuigongcishi as follows. \\
\begin{framed}
{\bf Algorithm 1} shousuigongcishi \\
/* this algorithm is mainly executed by magician and/or audience rather than computers. \\
Input: a sequential list of cards $A=(a, b, c, d, a, b, c, d)$ \\
Output: yes or no \\
{\bf Begin} \\
\hspace*{0.3cm} If your name has n1 words, then the first n1 cards of A is moved to the end of A, one by one. /* \textcircled{1} \\
\hspace*{0.3cm} The first three cards of A is bodily moved to anywhere of A, except for the begin and/or the end. /* \textcircled{2} \\
\hspace*{0.3cm} The first card is denoted as x1, and A:=tail(A). /* \textcircled{3} \\
\hspace*{0.3cm} /* tail(A) means a new list of cards consisting of the all cards of A except for the first one. \\
\hspace*{0.3cm} Select case your native place \\
\hspace*{0.6cm} [case southerner \\
\hspace*{0.9cm} [n2:=1]] \\
\hspace*{0.6cm} [case northerner \\
\hspace*{0.9cm} [n2:=2]] \\
\hspace*{0.6cm} [case unknown \\
\hspace*{0.9cm} [n2:=3]] \\
\hspace*{0.3cm} End select \\
\hspace*{0.3cm} The first n2 cards of A is bodily moved to anywhere of A, except for the begin and/or the end. /* \textcircled{4} \\
\hspace*{0.3cm} Select case your gender \\
\hspace*{0.6cm} [case male \\
\hspace*{0.9cm} [n3:=1]] \\
\hspace*{0.6cm} [case female \\
\hspace*{0.9cm} [n3:=2]] \\
\hspace*{0.3cm} End select \\
\hspace*{0.3cm} For k:= 1 to n3, do \\
\hspace*{0.6cm} A:=tail(A) \\
\hspace*{0.3cm} End for /* \textcircled{5} \\
\hspace*{0.3cm} For k:= 1 to 7, do \\
\hspace*{0.6cm} the first card of A is moved to the end of A \\
\hspace*{0.3cm} End for /* \textcircled{6} \\
\hspace*{0.3cm} For k:= 1 to 4, do \\
\hspace*{0.6cm} the first card of A is moved to the end of A \\
\hspace*{0.6cm} A:=tail(A) \\
\hspace*{0.3cm} End for /* \textcircled{7} \\
\hspace*{0.3cm} If your gender is male \\
\hspace*{0.6cm} the first card of A is moved to the end of A \\
\hspace*{0.6cm} A:=tail(A) \\
\hspace*{0.3cm} End if /* \textcircled{8} \\
\hspace*{0.3cm} If the only card in A = x1 \\
\hspace*{0.6cm} return "yes" \\
\hspace*{0.3cm} else \\
\hspace*{0.6cm} return "no" \\
\hspace*{0.3cm} End if /* \textcircled{9} \\
\noindent {\bf End} \\
\end{framed}
\centerline{\bf Algorithm 1}
{\bf Theorem 1} the algorithm 1 has a time complexity of O(n). \\
{\bf Proof} there are nine steps in this algorithm. And each step has a complexity of O(n1), O(3), O(2), O(n2), O(n3), O(7), O(8), O(2), O(1), respectively. Thus, the total time is a sum of the time of these steps, i.e., O(n). \, \, \, \, \, \, $\Box$ \\

\section{The Model Checking Algorithm} 
Does a formal model satisfy some temporal properties? This is magic model checking problem. On the one hand, an automaton can describe magic tricks. On the other hand, temporal logics such as linear temporal logic (LTL)[4], computational tree logic (CTL) [4] and interval temporal logic (ITL)[5], are employed to portray some temporal properties. Classical model checking algorithms [4][6] can be employed to verify whether these automata satisfy these properties. In addition, some personalized ways can also be considered for model checking some MTMs. For example, in terms of the algorithm for sousuigongcishi, algorithm 2 ensures the correctness of this magic design, while algorithm 3, algorithm 4, algorithm 5, and algorithm 6 ensure some temporal properties. It should be noted that the atomic proposition $p$ means the two half cards are matched with each other.
\begin{framed}
{\bf Algorithm 2} verify the correctness of shousuigongcishi \\
/* whether the last half card in hand match with the hidden half card or not? \\
/* this algorithm is executed by persons (especially for designers), or computers \\
Input: a sequential list of cards $A=(a, b, c, d, a, b, c, d)$ \\
Output: yes or no, true or false \\
{\bf Begin} \\
\hspace*{0.3cm} flag:=0; \\
\hspace*{0.3cm} For i=1:m \\
\hspace*{0.3cm} /* the combination of the values of n1, n2, n3, and anywhere in step 2 and 4, has num(paths)=m different values \\
\hspace*{0.6cm} If your name has n1 words, then the first n1 cards of A is moved to the end of A, one by one. /* \textcircled{1} \\
\hspace*{0.6cm} The first three cards of A is bodily moved to anywhere of A, except for the begin and/or the end. /* \textcircled{2} \\
\hspace*{0.6cm} The first card is denoted as x1, and A:=tail(A). /* \textcircled{3} \\
\hspace*{0.6cm} /* tail(A) means a new list of cards consisting of the all cards of A except for the first one. \\
\hspace*{0.6cm} /* from step 3 to step 9, we will model checking $AF(p \land empty)$, which is a ITL formula [5] \\
\hspace*{0.6cm} Select case your native place \\
\hspace*{0.9cm} [case southerner \\
\hspace*{1.2cm} [n2:=1]] \\
\hspace*{0.9cm} [case northerner \\
\hspace*{1.2cm} [n2:=2]] \\
\hspace*{0.9cm} [case unknown \\
\hspace*{1.2cm} [n2:=3]] \\
\hspace*{0.6cm} End select \\
\hspace*{0.6cm} The first n2 cards of A is bodily moved to anywhere of A, except for the begin and/or the end. /* \textcircled{4} \\
\hspace*{0.6cm} Select case your gender \\
\hspace*{0.9cm} [case male \\
\hspace*{1.2cm} [n3:=1]] \\
\hspace*{0.9cm} [case female \\
\hspace*{1.2cm} [n3:=2]] \\
\hspace*{0.6cm} End select \\
\hspace*{0.6cm} For k:= 1 to n3, do \\
\hspace*{0.9cm} A:=tail(A) \\
\hspace*{0.6cm} End for /* \textcircled{5} \\
\hspace*{0.6cm} For k:= 1 to 7, do \\
\hspace*{0.9cm} the first card of A is moved to the end of A \\
\hspace*{0.6cm} End for /* \textcircled{6} \\
\hspace*{0.6cm} For k:= 1 to 4, do \\
\hspace*{0.9cm} the first card of A is moved to the end of A \\
\hspace*{0.9cm} A:=tail(A) \\
\hspace*{0.6cm} End for /* \textcircled{7} \\
\hspace*{0.6cm} If your gender is male \\
\hspace*{0.9cm} the first card of A is moved to the end of A \\
\hspace*{0.9cm} A:=tail(A) \\
\hspace*{0.6cm} End if /* \textcircled{8} \\
\hspace*{0.6cm} If the only card in A = x1 \\
\hspace*{0.9cm} return "yes" \\
\hspace*{0.9cm} flag:=flag+1; \\
\hspace*{0.6cm} else \\
\hspace*{0.9cm} return "no" \\
\hspace*{0.6cm} End if /* \textcircled{9} \\
\hspace*{0.3cm} End for \\
\hspace*{0.3cm} If flag=m, then MC result of $AF(p \land empty)$ is "true", else MC result of $AF(p \land empty)$ is "false" \\
\noindent {\bf End} \\
\end{framed}
\centerline{\bf Algorithm 2}
\noindent {\bf Theorem 2} the algorithm 2 has a time complexity of O(m*n). \\
{\bf Proof} The algorithm has a loop. The loop is executed m times. There are nine steps in the loop body. Each loop consumes the following time: O(n1)+O(3)+O(2)+O(n2)+O(n3)+O(7)+O(8)+O(2)+O(2)=O(n). Thus, the total time is O(m*n). \, \, \, \, \, \, $\Box$ \\
\begin{framed}
{\bf Algorithm 3} MC AFp for shousuigongcishi \\
/* whether the last half card in the sequence in hand is equal to the hidden half one in one step (after step 3) for every playing. \\
/* this algorithm is executed by persons (especially for designers), or computers \\
Input: a sequential list of cards $A=(a, b, c, d, a, b, c, d)$ \\
Output: yes or no, true or false \\
{\bf Begin} \\
\hspace*{0.3cm} flag:=0; \\
\hspace*{0.3cm} For i=1:m \\
\hspace*{0.3cm} /* the combination of the values of n1, n2, n3, and anywhere in step 2 and 4, has num(paths)=m different values \\
\hspace*{0.6cm} If your name has n1 words, then the first n1 cards of A is moved to the end of A, one by one. /* \textcircled{1} \\
\hspace*{0.6cm} The first three cards of A is bodily moved to anywhere of A, except for the begin and/or the end. /* \textcircled{2} \\
\hspace*{0.6cm} The first card is denoted as x1, and A:=tail(A). /* \textcircled{3} \\
\hspace*{0.6cm} /* tail(A) means a new list of cards consisting of the all cards of A except for the first one. \\
\hspace*{0.6cm} /* from step 3 to step 9, we will model checking $AFp$, which is a CTL formula [4] \\
\hspace*{0.6cm} Select case your native place \\
\hspace*{0.9cm} [case southerner \\
\hspace*{1.2cm} [n2:=1]] \\
\hspace*{0.9cm} [case northerner \\
\hspace*{1.2cm} [n2:=2]] \\
\hspace*{0.9cm} [case unknown \\
\hspace*{1.2cm} [n2:=3]] \\
\hspace*{0.6cm} End select \\
\hspace*{0.6cm} The first n2 cards of A is bodily moved to anywhere of A, except for the begin and/or the end. \\
\hspace*{0.6cm} If last card of A is equal to x1, then flag:=1 /* \textcircled{4} \\
\hspace*{0.6cm} Select case your gender \\
\hspace*{0.9cm} [case male \\
\hspace*{1.2cm} [n3:=1]] \\
\hspace*{0.9cm} [case female \\
\hspace*{1.2cm} [n3:=2]] \\
\hspace*{0.6cm} End select \\
\hspace*{0.6cm} For k:= 1 to n3, do \\
\hspace*{0.9cm} A:=tail(A) \\
\hspace*{0.6cm} End for \\
\hspace*{0.6cm} If last card of A is equal to x1, then flag:=1 /* \textcircled{5} \\
\hspace*{0.6cm} For k:= 1 to 7, do \\
\hspace*{0.9cm} the first card of A is moved to the end of A \\
\hspace*{0.6cm} End for \\
\hspace*{0.6cm} If last card of A is equal to x1, then flag:=1 /* \textcircled{6} \\
\hspace*{0.6cm} For k:= 1 to 4, do \\
\hspace*{0.9cm} the first card of A is moved to the end of A \\
\hspace*{0.9cm} A:=tail(A) \\
\hspace*{0.6cm} End for \\
\hspace*{0.6cm} If last card of A is equal to x1, then flag:=1 /* \textcircled{7} \\
\hspace*{0.6cm} If your gender is male \\
\hspace*{0.9cm} the first card of A is moved to the end of A \\
\hspace*{0.9cm} A:=tail(A) \\
\hspace*{0.6cm} End if \\
\hspace*{0.6cm} If last card of A is equal to x1, then flag:=1 /* \textcircled{8} \\
\hspace*{0.6cm} If the only card in A = x1 \\
\hspace*{0.9cm} return "yes" \\
\hspace*{0.9cm} flag:=1; \\
\hspace*{0.6cm} else \\
\hspace*{0.9cm} return "no" \\
\hspace*{0.6cm} End if /* \textcircled{9} \\
\hspace*{0.3cm} End for \\
\hspace*{0.3cm} flag:=flag+1; \\
\hspace*{0.3cm} If flag=m, then MC result of $AFp$ is "true", else MC result of $AFp$ is "false" \\
\noindent {\bf End} \\
\end{framed}
\centerline{\bf Algorithm 3}
\noindent {\bf Theorem 3} the algorithm 3 has a time complexity of O(m*n). \\
{\bf Proof} This is the similar way with the process of proving theorem 2. \, \, \, \, \, \, $\Box$ \\
\begin{framed}
{\bf Algorithm 4} MC AGp for shousuigongcishi \\
/* whether the last half card in the sequence in hand is equal to the hidden half one in each step (after step 3) for every playing. \\
/* this algorithm is executed by persons (especially for designers), or computers \\
Input: a sequential list of cards $A=(a, b, c, d, a, b, c, d)$ \\
Output: yes or no, true or false \\
{\bf Begin} \\
\hspace*{0.3cm} flag:=0; \\
\hspace*{0.3cm} For i=1:m \\
\hspace*{0.3cm} /* the combination of the values of n1, n2, n3, and anywhere in step 2 and 4, has num(paths)=m different values \\
\hspace*{0.6cm} If your name has n1 words, then the first n1 cards of A is moved to the end of A, one by one. /* \textcircled{1} \\
\hspace*{0.6cm} The first three cards of A is bodily moved to anywhere of A, except for the begin and/or the end. /* \textcircled{2} \\
\hspace*{0.6cm} The first card is denoted as x1, and A:=tail(A). /* \textcircled{3} \\
\hspace*{0.6cm} /* tail(A) means a new list of cards consisting of the all cards of A except for the first one. \\
\hspace*{0.6cm} /* from step 3 to step 9, we will model checking $AGp$, which is a CTL formula [4] \\
\hspace*{0.6cm} Select case your native place \\
\hspace*{0.9cm} [case southerner \\
\hspace*{1.2cm} [n2:=1]] \\
\hspace*{0.9cm} [case northerner \\
\hspace*{1.2cm} [n2:=2]] \\
\hspace*{0.9cm} [case unknown \\
\hspace*{1.2cm} [n2:=3]] \\
\hspace*{0.6cm} End select \\
\hspace*{0.6cm} The first n2 cards of A is bodily moved to anywhere of A, except for the begin and/or the end. \\
\hspace*{0.6cm} If last card of A is equal to x1, then flag:=flag+1 /* \textcircled{4} \\
\hspace*{0.6cm} Select case your gender \\
\hspace*{0.9cm} [case male \\
\hspace*{1.2cm} [n3:=1]] \\
\hspace*{0.9cm} [case female \\
\hspace*{1.2cm} [n3:=2]] \\
\hspace*{0.6cm} End select \\
\hspace*{0.6cm} For k:= 1 to n3, do \\
\hspace*{0.9cm} A:=tail(A) \\
\hspace*{0.6cm} End for \\
\hspace*{0.6cm} If last card of A is equal to x1, then flag:=flag+1 /* \textcircled{5} \\
\hspace*{0.6cm} For k:= 1 to 7, do \\
\hspace*{0.9cm} the first card of A is moved to the end of A \\
\hspace*{0.6cm} End for \\
\hspace*{0.6cm} If last card of A is equal to x1, then flag:=flag+1 /* \textcircled{6} \\
\hspace*{0.6cm} For k:= 1 to 4, do \\
\hspace*{0.9cm} the first card of A is moved to the end of A \\
\hspace*{0.9cm} A:=tail(A) \\
\hspace*{0.6cm} End for \\
\hspace*{0.6cm} If last card of A is equal to x1, then flag:=flag+1 /* \textcircled{7} \\
\hspace*{0.6cm} If your gender is male \\
\hspace*{0.9cm} the first card of A is moved to the end of A \\
\hspace*{0.9cm} A:=tail(A) \\
\hspace*{0.6cm} End if \\
\hspace*{0.6cm} If last card of A is equal to x1, then flag:=flag+1 /* \textcircled{8} \\
\hspace*{0.6cm} If the only card in A = x1 \\
\hspace*{0.9cm} return "yes" \\
\hspace*{0.9cm} flag:=flag+1; \\
\hspace*{0.6cm} else \\
\hspace*{0.9cm} return "no" \\
\hspace*{0.6cm} End if /* \textcircled{9} \\
\hspace*{0.3cm} End for \\
\hspace*{0.3cm} If flag=6*m, then MC result of $AGp$ is "true", else MC result of $AGp$ is "false" \\
\noindent {\bf End} \\
\end{framed}
\centerline{\bf Algorithm 4}
\noindent {\bf Theorem 4} the algorithm 4 has a time complexity of O(m*n). \\
{\bf Proof} This is the similar way with the process of proving theorem 3. \, \, \, \, \, \, $\Box$ \\
\begin{framed}
{\bf Algorithm 5} MC EFp for shousuigongcishi \\
/* whether the last half card in the sequence in hand is equal to the hidden half one in one step (after step 3) for one playing. \\
/* this algorithm is executed by persons (especially for designers), or computers \\
Input: a sequential list of cards $A=(a, b, c, d, a, b, c, d)$ \\
Output: yes or no, true or false \\
{\bf Begin} \\
\hspace*{0.3cm} flag:=0; \\
\hspace*{0.3cm} For i=1:m \\
\hspace*{0.3cm} /* the combination of the values of n1, n2, n3, and anywhere in step 2 and 4, has num(paths)=m different values \\
\hspace*{0.6cm} If your name has n1 words, then the first n1 cards of A is moved to the end of A, one by one. /* \textcircled{1} \\
\hspace*{0.6cm} The first three cards of A is bodily moved to anywhere of A, except for the begin and/or the end. /* \textcircled{2} \\
\hspace*{0.6cm} The first card is denoted as x1, and A:=tail(A). /* \textcircled{3} \\
\hspace*{0.6cm} /* tail(A) means a new list of cards consisting of the all cards of A except for the first one. \\
\hspace*{0.6cm} /* from step 3 to step 9, we will model checking $EFp$, which is a CTL formula [4] \\
\hspace*{0.6cm} Select case your native place \\
\hspace*{0.9cm} [case southerner \\
\hspace*{1.2cm} [n2:=1]] \\
\hspace*{0.9cm} [case northerner \\
\hspace*{1.2cm} [n2:=2]] \\
\hspace*{0.9cm} [case unknown \\
\hspace*{1.2cm} [n2:=3]] \\
\hspace*{0.6cm} End select \\
\hspace*{0.6cm} The first n2 cards of A is bodily moved to anywhere of A, except for the begin and/or the end. \\
\hspace*{0.6cm} If last card of A is equal to x1, then flag:=1 /* \textcircled{4} \\
\hspace*{0.6cm} Select case your gender \\
\hspace*{0.9cm} [case male \\
\hspace*{1.2cm} [n3:=1]] \\
\hspace*{0.9cm} [case female \\
\hspace*{1.2cm} [n3:=2]] \\
\hspace*{0.6cm} End select \\
\hspace*{0.6cm} For k:= 1 to n3, do \\
\hspace*{0.9cm} A:=tail(A) \\
\hspace*{0.6cm} End for \\
\hspace*{0.6cm} If last card of A is equal to x1, then flag:=1 /* \textcircled{5} \\
\hspace*{0.6cm} For k:= 1 to 7, do \\
\hspace*{0.9cm} the first card of A is moved to the end of A \\
\hspace*{0.6cm} End for \\
\hspace*{0.6cm} If last card of A is equal to x1, then flag:=1 /* \textcircled{6} \\
\hspace*{0.6cm} For k:= 1 to 4, do \\
\hspace*{0.9cm} the first card of A is moved to the end of A \\
\hspace*{0.9cm} A:=tail(A) \\
\hspace*{0.6cm} End for \\
\hspace*{0.6cm} If last card of A is equal to x1, then flag:=1 /* \textcircled{7} \\
\hspace*{0.6cm} If your gender is male \\
\hspace*{0.9cm} the first card of A is moved to the end of A \\
\hspace*{0.9cm} A:=tail(A) \\
\hspace*{0.6cm} End if \\
\hspace*{0.6cm} If last card of A is equal to x1, then flag:=1 /* \textcircled{8} \\
\hspace*{0.6cm} If the only card in A = x1 \\
\hspace*{0.9cm} return "yes" \\
\hspace*{0.9cm} flag:=1; \\
\hspace*{0.6cm} else \\
\hspace*{0.9cm} return "no" \\
\hspace*{0.6cm} End if /* \textcircled{9} \\
\hspace*{0.3cm} End for \\
\hspace*{0.3cm} flag:=flag+1; \\
\hspace*{0.3cm} If flag$>$0, then MC result of $EFp$ is "true", else MC result of $EFp$ is "false" \\
\noindent {\bf End} \\
\end{framed}
\centerline{\bf Algorithm 5}
\noindent {\bf Theorem 5} the algorithm 5 has a time complexity of O(m*n). \\
{\bf Proof} This is the similar way with the process of proving theorem 4. \, \, \, \, \, \, $\Box$ \\
\begin{framed}
{\bf Algorithm 6} MC EGp for shousuigongcishi \\
/* whether the last half card in the sequence in hand is equal to the hidden half one in each step (after step 3) for one playing. \\
/* this algorithm is executed by persons (especially for designers), or computers \\
Input: a sequential list of cards $A=(a, b, c, d, a, b, c, d)$ \\
Output: yes or no, true or false \\
{\bf Begin} \\
\hspace*{0.3cm} flag2:=0; \\
\hspace*{0.3cm} For i=1:m \\
\hspace*{0.3cm} /* the combination of the values of n1, n2, n3, and anywhere in step 2 and 4, has num(paths)=m different values \\
\hspace*{0.6cm} flag1:=0; \\
\hspace*{0.6cm} If your name has n1 words, then the first n1 cards of A is moved to the end of A, one by one. /* \textcircled{1} \\
\hspace*{0.6cm} The first three cards of A is bodily moved to anywhere of A, except for the begin and/or the end. /* \textcircled{2} \\
\hspace*{0.6cm} The first card is denoted as x1, and A:=tail(A). /* \textcircled{3} \\
\hspace*{0.6cm} /* tail(A) means a new list of cards consisting of the all cards of A except for the first one. \\
\hspace*{0.6cm} /* from step 3 to step 9, we will model checking $EGp$, which is a CTL formula [4] \\
\hspace*{0.6cm} Select case your native place \\
\hspace*{0.9cm} [case southerner \\
\hspace*{1.2cm} [n2:=1]] \\
\hspace*{0.9cm} [case northerner \\
\hspace*{1.2cm} [n2:=2]] \\
\hspace*{0.9cm} [case unknown \\
\hspace*{1.2cm} [n2:=3]] \\
\hspace*{0.6cm} End select \\
\hspace*{0.6cm} The first n2 cards of A is bodily moved to anywhere of A, except for the begin and/or the end. \\
\hspace*{0.6cm} If last card of A is equal to x1, then flag1:=flag1+1 /* \textcircled{4} \\
\hspace*{0.6cm} Select case your gender \\
\hspace*{0.9cm} [case male \\
\hspace*{1.2cm} [n3:=1]] \\
\hspace*{0.9cm} [case female \\
\hspace*{1.2cm} [n3:=2]] \\
\hspace*{0.6cm} End select \\
\hspace*{0.6cm} For k:= 1 to n3, do \\
\hspace*{0.9cm} A:=tail(A) \\
\hspace*{0.6cm} End for \\
\hspace*{0.6cm} If last card of A is equal to x1, then flag1:=flag1+1 /* \textcircled{5} \\
\hspace*{0.6cm} For k:= 1 to 7, do \\
\hspace*{0.9cm} the first card of A is moved to the end of A \\
\hspace*{0.6cm} End for \\
\hspace*{0.6cm} If last card of A is equal to x1, then flag1:=flag1+1 /* \textcircled{6} \\
\hspace*{0.6cm} For k:= 1 to 4, do \\
\hspace*{0.9cm} the first card of A is moved to the end of A \\
\hspace*{0.9cm} A:=tail(A) \\
\hspace*{0.6cm} End for \\
\hspace*{0.6cm} If last card of A is equal to x1, then flag1:=flag1+1 /* \textcircled{7} \\
\hspace*{0.6cm} If your gender is male \\
\hspace*{0.9cm} the first card of A is moved to the end of A \\
\hspace*{0.9cm} A:=tail(A) \\
\hspace*{0.6cm} End if \\
\hspace*{0.6cm} If last card of A is equal to x1, then flag1:=flag1+1 /* \textcircled{8} \\
\hspace*{0.6cm} If the only card in A = x1 \\
\hspace*{0.9cm} return "yes" \\
\hspace*{0.9cm} flag1:=flag1+1; \\
\hspace*{0.6cm} else \\
\hspace*{0.9cm} return "no" \\
\hspace*{0.6cm} End if /* \textcircled{9} \\
\hspace*{0.6cm} If flag1=6 then flag2:=1 \\
\hspace*{0.3cm} End for \\
\hspace*{0.3cm} flag:=flag+1; \\
\hspace*{0.3cm} If flag2=1, then MC result of $EGp$ is "true", else MC result of $EGp$ is "false" \\
\noindent {\bf End} \\
\end{framed}
\centerline{\bf Algorithm 6}
\noindent {\bf Theorem 6} the algorithm 6 has a time complexity of O(m*n). \\
{\bf Proof} This is the similar way with the process of proving theorem 5. \, \, \, \, \, \, $\Box$ \\

\section{Experiments}
\subsection{Experimental Platform}
\begin{itemize}
\item CPU: i9 10900, 2.80 Ghz;
\item Memory: 32GB; 
\item OS: windows 10;
\item Tool: Matlab 2017.
\end{itemize}

\subsection{Experimental Procedure}
(1) algorithm 2 is encoded via matlab, before running and getting the running results of shousuigongcishi, MC results of $AF(p \land empty)$ and running time. \\
(2) algorithm 3 is encoded via matlab, before running and getting MC results of $AFp$ and time. \\
(3) algorithm 4 is encoded via matlab, before running and getting MC results of $AGp$ and time. \\
(4) algorithm 5 is encoded via matlab, before running and getting MC results of $EFp$ and time. \\
(5) algorithm 6 is encoded via matlab, before running and getting MC results of $EGp$ and time. \\

\subsection{Experimental Results}
As shown in fig.1 and fig.2, we get some experimental results. \\
 
     \begin{figure}
		\centering
			\includegraphics[scale=0.8]{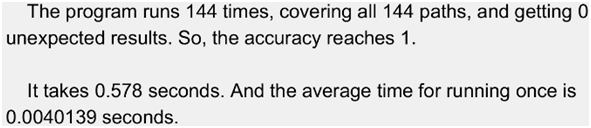}
			\caption{MC results of $AF(p \land empty)$ and running time}
			\label{fig1}
	\end{figure}

	\begin{figure}
		\centering
		\subfigure[AF]{
			\begin{minipage}[b]{0.8\textwidth}
				\centering \includegraphics[scale=0.8]{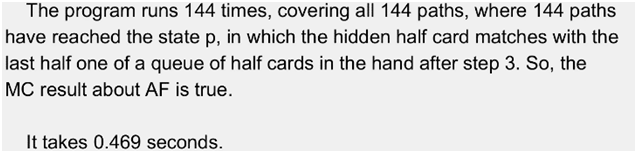} \\
                     \centering \includegraphics[scale=0.8]{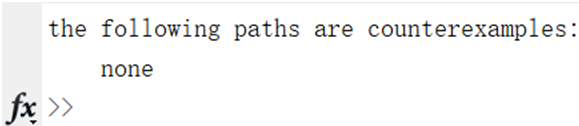} \\
				\label{fig2-1}
			\end{minipage}
		}
		
		\subfigure[AG]
		{
			\begin{minipage}[b]{0.8\textwidth}
				\centering \includegraphics[scale=0.8]{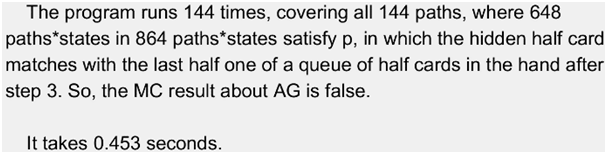} \\
                     \centering \includegraphics[scale=0.8]{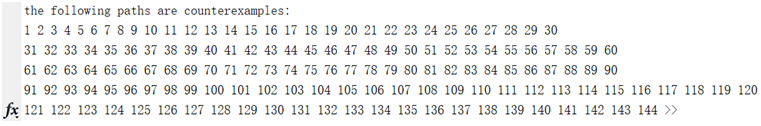} \\
				\label{fig2-2}
			\end{minipage}
		}
		
		\subfigure[EF]{
			\begin{minipage}[c]{0.8\textwidth}
				\centering \includegraphics[scale=0.8]{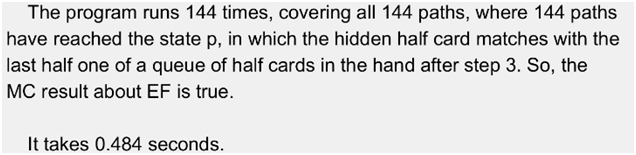} \\
                     \centering \includegraphics[scale=0.8]{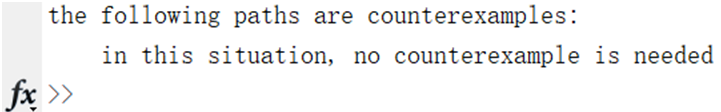} \\
				\label{fig2-3}
			\end{minipage}
		}

          \subfigure[EG]{
			\begin{minipage}[d]{0.8\textwidth}
				\centering \includegraphics[scale=0.8]{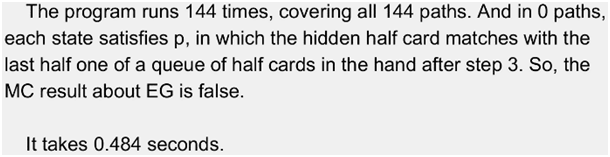} \\
                     \centering \includegraphics[scale=0.8]{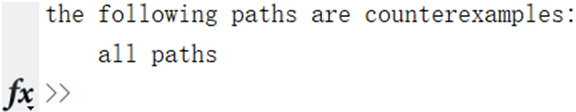} \\
				\label{fig2-4}
			\end{minipage}
		}
		
		\caption{MC results}
		\label{fig2}
	\end{figure}

\section{Discussions}

     \begin{table}  
		\centering		
		\caption{some differences}
		\label{tab1}
		
		\begin{tabular}{|c|c|c|} 
			\hline  
			{\bf Formal mechanisms} & {\bf Presented in this study} & {\bf The classical ones} \\
			\hline  
			Automata / TM  & \multirow{3}{6cm}{Executed by persons/computers, and modeling/verifying some magicians' actions} & \multirow{3}{5.6cm}{Executed by computers / modeling/verifying some computers' actions} \\
			\cline{1-1}
			Algorithms  &  &   \\
			\cline{1-1}
			Model checking  &  &  \\
			\hline
		\end{tabular}  
	\end{table}

In short, table 1 can help us understanding the proposed models and methods. In addition, the proposed models and methods have the similar expressing forms in terms of syntax and semantics with the classical ones. It is clear that, the experiments in the previous section can also be executed by human rather than computers. \\
\hspace*{0.4cm} Technically speaking / from a computer science standpoint / perhaps in a manner of speaking, Mr. Liu called on countless audience to "execute the algorithm" parallelly at that time [1][2]. Theoretically, he only needs 144 viewers and six minutes so that they can check all 144 paths to verify the correctness of the logic design of this magic trick, supposing each audience's name has two or three words. More importantly, many people derived pleasure in the process of this magic show, at that time. \\
\hspace*{0.4cm} It should be noted that, a magic algorithm maybe has some unnecessary steps, if we only need computational results. For example, shousuigongcishi's computational function is finding two matched half cards. We can archive this goal before step 3. However, the following "redundant" steps make the thing more interesting. After all, it is a magic trick. If time complexity is very vital in traditional computing theory, entertainment rather than time is more vital in "computing" theory of computational magic tricks. \\

\section{Conclusion}
In this study, some formal concept about magic formal models, magic algorithms and magic model checking are presented, and they are formal mechanisms for magicians' logic actions. In this case, there is another method, i.e., theorem proving, in verifying the correctness of this CM. Whatever theorem proving or model checking, formal verification and computational magic can be expected to help architects in the process of designing some interesting, attractive and wonderful logic/computational tricks, especially for more complex, stronger and more safe running large scale CM.

\end{document}